\documentclass[aps,twocolumn]{revtex4-1}

\usepackage{amsmath}  
\usepackage{amsfonts} 
\usepackage{graphicx, subfigure} 

\usepackage{color}
\definecolor{r}{rgb}{1,0,0}

\usepackage{epsfig}
\definecolor{g}{rgb}{0,1,0}

\definecolor{b}{rgb}{0,0,1}

\definecolor{m}{rgb}{1,0,1}

\def\picwidth{0.9\columnwidth}

\begin{document}

\title{Free Cooling of a Granular Gas in Three Dimensions}

\author{Kirsten Harth, Torsten Trittel, Sandra Wegner, and
Ralf Stannarius }
\affiliation{Institute for Experimental Physics, Otto von Guericke University, Magdeburg, Germany}

\date{\today}

\begin{abstract}
Granular gases as dilute ensembles of particles in random motion are not only at the basis of elementary structure-forming processes in the universe and involved in many industrial and natural phenomena,
but also excellent models to study fundamental statistical dynamics.
A vast number of theoretical and
numerical investigations have dealt with this apparently simple non-equilibrium system.
The essential difference to molecular gases
is the energy dissipation in particle collisions, a subtle distinction with immense impact on their
global dynamics. Its most striking manifestation is the so-called granular cooling,
the gradual loss of mechanical energy in absence of external excitation.

We report an experimental study of homogeneous cooling of three-dimensional (3D) granular gases
in microgravity.
Surprisingly, the asymptotic scaling $E(t)\propto t^{-2}$ obtained by Haff's minimal model
[J. Fluid Mech. {\bf 134} 401 (1983)]
proves to be robust, despite the violation of several of its central assumptions. The shape anisotropy of the grains influences the characteristic time of energy loss quantitatively, but not qualitatively.
We compare kinetic energies in the individual degrees of freedom, and find a slight predominance of the translational motions.
In addition, we detect a certain preference of the grains to align with their long axis in flight direction, a feature known from active matter or animal flocks, and the onset of clustering.
\end{abstract}

\maketitle

\section{Granular gases}

Cars and pedestrians in traffic, migrating groups of animals or bacteria,
bubbles in fluid flows, ice crystals above snow avalanches, grains in sand storms are examples
of large ensembles of particles in which occasional interactions of individual constituents govern the
emergence of collective dynamic patterns. All these are inherently out of thermal equilibrium.
Granular gases, i.e. dilute ensembles of grains interacting by dissipative
collisions, represent the simplest of such many-body systems, without long-range interactions.
At first glance, they resemble analogues of molecular gases.
The dissipative character of particle interactions, however, alters the ensemble properties fundamentally:
Clustering
(e. g. \cite{Goldhirsch1993,Falcon1999,Sapozhnikov2003,Opsomer2011,Herminghaus2017,Hummel2016,Hummel2016b,%
Kudrolli1997,Maass2008}),
non-Gaussian velocity
distributions (e. g. \cite{Nichol2012,Hou2008,Olafsen1998,Olafsen1999,Tatsumi2009,Losert1999,Kudrolli2000,Schmick2008,%
Kohlstedt2005,Aranson2002,Rouyer2000,%
Losert1999,Huan2004,Tatsumi2009,Costantini2005,Wildman2009,Harth2013,Scholz2017}), and anomalous scaling of the pressure
\cite{Evesque2001,Geminard2004,Falcon2006} are but a few documented examples.

Most prominent is the permanent loss of kinetic energy in absence of external forcing,
called granular cooling \cite{Haff1983}: Starting from an initially excited state with spatially
homogeneous statistical properties, the ensemble enters an initial period of homogeneous energy loss.
At longer time scales, the grains can spontaneously cluster.
Such granular clustering is a key ingredient for the formation of planetesimals and larger objects
in solar systems~\cite{Wurm1998,Johansen2007,Williams2011}.

But even today, fundamental features of such ensembles are only little understood, quantitative experiments are very much needed.
Analytical and numerical studies in the past 20 years produced results strongly depending on the
simplifications made and assumptions of specific grain properties (see, e. g. \cite{Goldhirsch1993,Kanzaki2010,Luding1998,Villemot2012,Rubio-Largo2016}).
Often, spherical grains under ideal initial and boundary conditions were considered.
Experiments were focused on quasi-2D layers
\cite{Olafsen1998,Olafsen1999,Kudrolli1997,Maass2008,Daniels2009,Nichol2012,Tatsumi2009,Hou2008,%
Feitosa2002,Losert1999,Leconte2006,Burton2013}, where, in addition to the above-mentioned features,
the equipartition of energies was tested \cite{Feitosa2002,Daniels2009,Nichol2012},
collisions of particles \cite{Leconte2006} and clusters \cite{Maass2008,Burton2013} were analyzed.

Preparation of a freely cooling granular gas in 3D is challenging, it is practically
impossible under normal gravity.
Sounding rockets, satellites and drop towers offer
appropriate conditions \cite{Harth2015}: excellent micro-gravity ($\mu$g) quality,
down to residual accelerations of $10^{-5}$~m/s$^2$.
In a pioneering $\mu$g experiment, dynamical clustering of monodisperse spheres was reported by Falcon et al.~\cite{Falcon1999}, but a  quantitative analysis at the grain-level was not possible for technical limitations.
{\em Rod-shaped} grains offer experimental advantages over spheres: a much shorter mean free path at comparable filling fractions~\cite{Harth2013,Harth2017}, a more random energy injection by vibrating container walls~\cite{Wright2006,Trittel2017}, and an efficient energy redistribution among all
degrees of freedom (DoF) in collisions~\cite{Rubio-Largo2016}.
The latter two features can reduce spatial inhomogeneities.
Translations and rotations can be followed in 3D~\cite{Harth2013b}.
We present results of the first experimental investigation of a homogeneously cooling 3D granular gas.

\section {Experiment}
\begin{figure*}[ht]
   \includegraphics[width=0.8\textwidth]{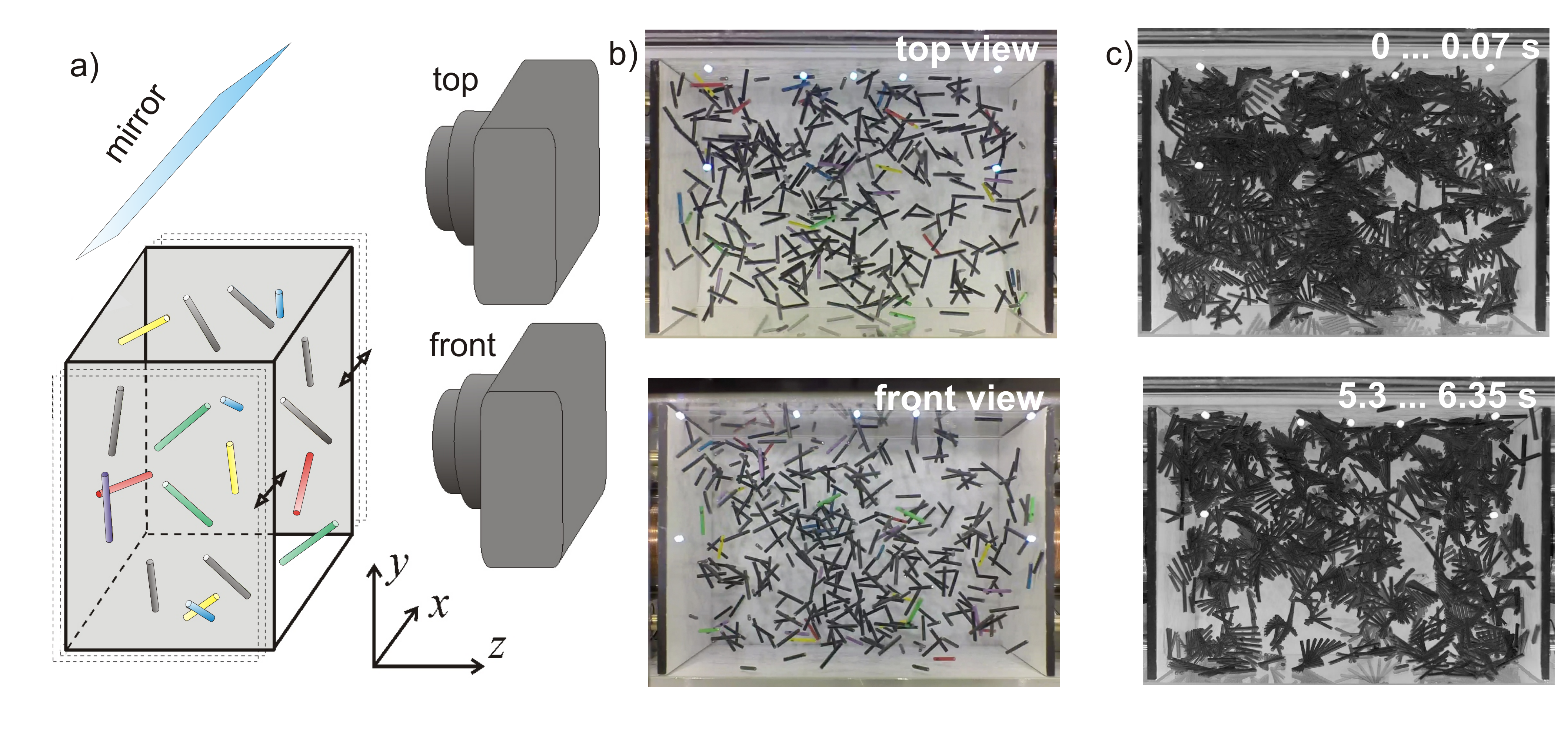}\\
  \caption{Experimental setup for microgravity experiments with granular gases: a) Sketch of the experimental setup and definition of the coordinate system.
  The two side walls can be vibrated mechanically, the top and front walls are transparent. b) Typical frames
  of the top and front camera videos. Colored particles are tracked, black rods provide a thermal background. c) eight superimposed frames of the top view camera taken at the beginning of the cooling experiment (homogeneous state), and at the end of the experiment (slight clustering). The frame rate is scaled with the mean velocity during the respective period.}
  \label{Fig:ExperimentBasics}
\end{figure*}

Ensembles of 374 rods
of $\ell=10$~mm length and $d=1.35$~mm diameter are studied in a container
of 11.2 cm $\times$ 8.0 cm $\times$ 8.0 cm (Fig.~\ref{Fig:ExperimentBasics} a,b) during  $\approx 9$~s of micro-gravity realized in the ZARM Drop Tower, Bremen. The corresponding volume fraction of grains
is $\phi=0.75~\%$. The mean free path estimated from the filling fraction and the rod dimensions is
$\lambda \approx \sqrt{2}d^2/[ \phi  (\ell + 7.55 d + 2.02d^2/\ell)]
\approx 1.65~\rm cm$, well below the Knudsen regime ($\lambda > $ container size). The rods are custom-made from insulated copper wire, their mass is $m=37.5$~mg,
moments of inertia for rotations around the rod axis
and perpendicular to it are $J_{\parallel}= 4.6$~pN\,m and
$J_{\perp}=315$~pN\,m, respectively.

\begin{figure}[ht]
   a)\includegraphics[width=\picwidth]{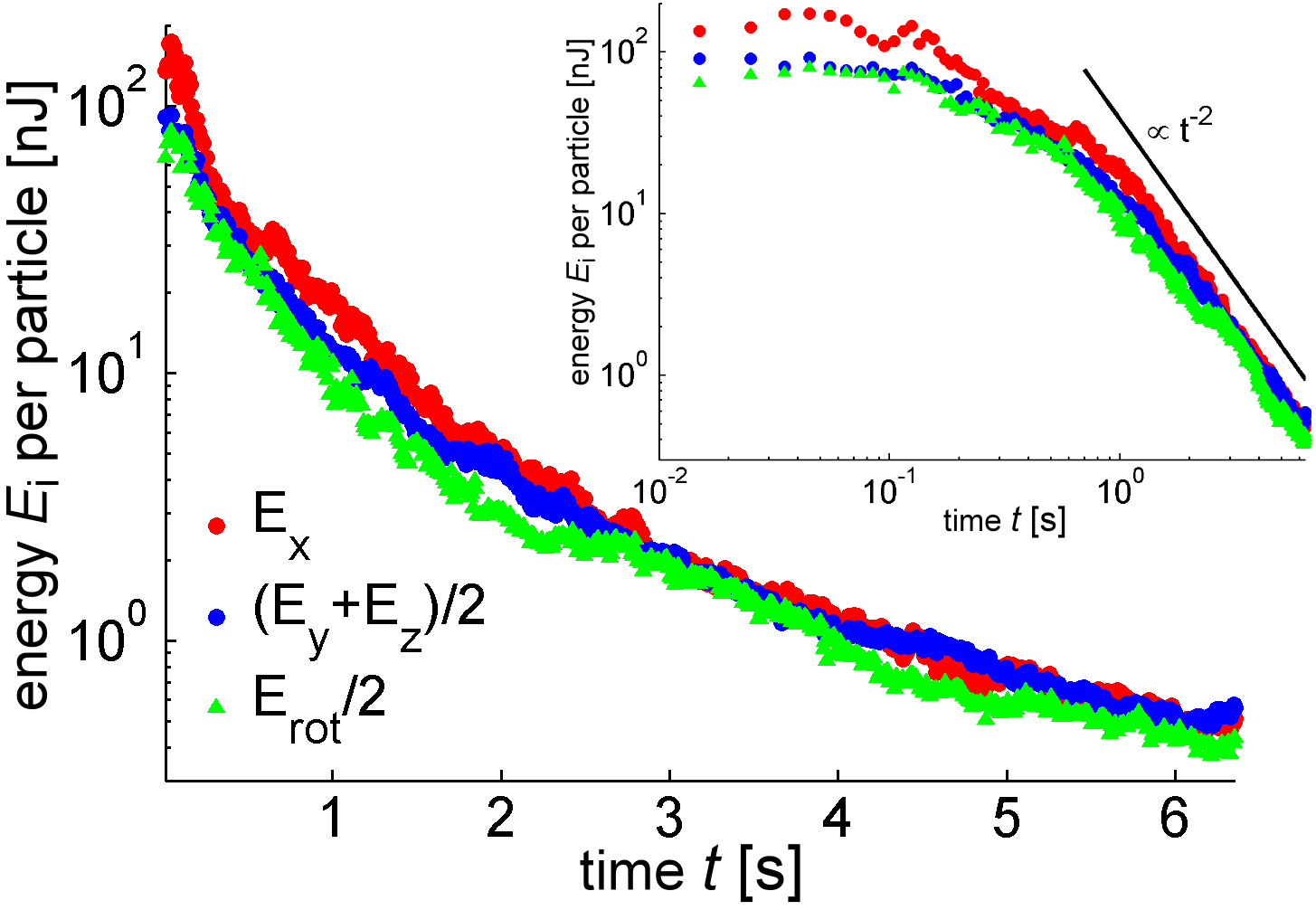}\\
   b)\includegraphics[width=\picwidth]{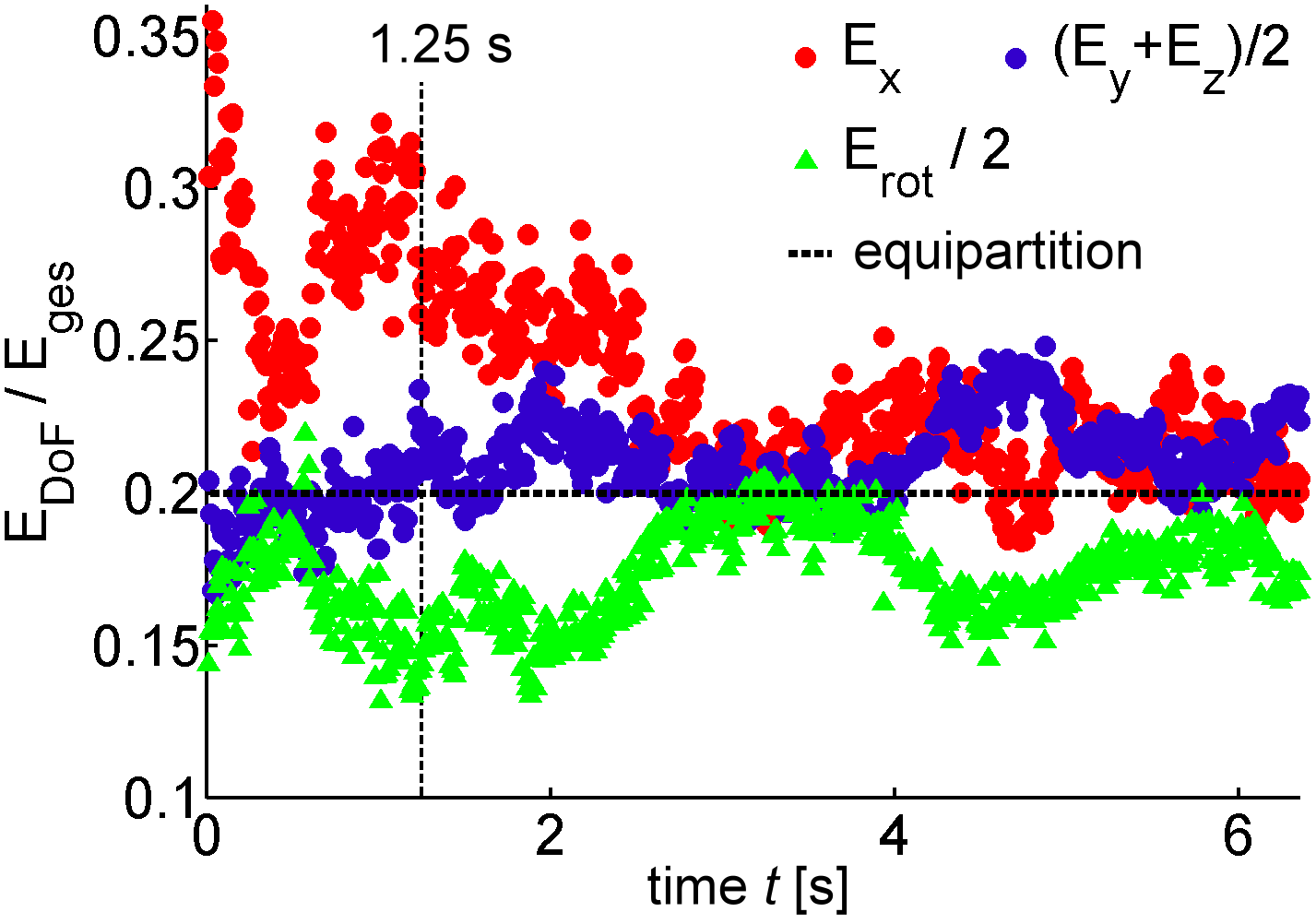}
  \caption{The energy is partitioned during granular cooling: a) Decay of the kinetic energy per particle in the individual degrees of freedom:
  after $\approx$ 2.5 s, the energy partition is steady within statistical fluctuations, and the decay proceeds as $t^{-2}$ as predicted by Eq. (\ref{eq:Haff}) (see logarithmic plot in the inset).
  b) Energies in the different types of DoF, all translational DoF equilibrate within statistical fluctuations, rotations about the short rod axes are systematically weaker excited by about 10-20\%, the rotations about the long axis were not considered here.
  }
  \label{Fig:2}
\end{figure}

\paragraph{Steady excitation state:} During the initial 2 seconds of microgravity, the grain ensemble is excited mechanically.
In the initial, driven state, one finds an excess of
translational energy in the direction of excitation $x$ (normal to the vibrating walls),
$\langle E_x(0)\rangle\approx135~\rm nJ$ per grain.
Rotations and translations in the directions $y$ and $z$ are only weakly excited by the vibrating walls \cite{Trittel2017}, they are driven via rod-rod collisions.
The indirectly excited spatial directions have equal average energies, initially $\langle E_{y}+E_{z}\rangle /2\approx 90~\rm nJ$, and
$\langle E_{\rm rot}/2\rangle \approx 64~\rm nJ$ for rotations about the short rod axes (these two rotations cannot be distinguished in our experiment, we can only determine their sum $E_{\rm rot}$). The violation of equipartition in the driven state has been described earlier~\cite{Harth2013,Harth2013b}.
After the excitation is stopped ($t_0=0$), videos are recorded and a particle-based statistical analysis (see Methods) is performed to evaluate the evolution of the ensemble dynamics.
Of primary interest is the energy partition and energy loss. The total kinetic energy decays by almost 3 orders of magnitude during the observation period of about 7~s, as seen in Fig.~\ref{Fig:2}.

\paragraph{Spatial homogeneity during cooling:}
Throughout the cooling process we found only a marginal tendency of clustering. This is difficult to evaluate quantitatively but visualized best by overlaying image sequences from a single experiment. Figure
\ref{Fig:ExperimentBasics}c shows two such typical overlays. The upper one was recorded immediately after $t_0$, eight subsequent frames of the top camera video were overlayed, the field of view is more or less
uniformly covered by rods. The bottom view is a superposition of eight frames from the final phase of the
experiment. Since the mean velocity is slower by a factor of $\approx 15$, every 15th frame was chosen for the overlay. One can recognize some inhomogeneities in the particle distribution, but the effect is marginal.

\paragraph{Partition of the kinetic energy:}
After the excitation stopped, the kinetic energy is gradually redistributed by collisions.
Thereby, the partition among the DoF changes drastically, as shown in Fig.~\ref{Fig:2}, top.
Within statistical fluctuations, a steady distribution of the kinetic
energy among all DoF is reached after $\approx 2 \dots 2.5~\rm s$. The initial
excess of $E_x$ has vanished. However, the kinetic energies of
rotations around the short rod axes remain
slightly smaller than those of the translational DoF, as seen in Fig.~\ref{Fig:2}.
These results are in qualitative agreement with simulations of frictionless
ellipsoids~\cite{Villemot2012} and rods~\cite{Rubio-Largo2016} that predicted a small excess of
translational over rotational energies per DoF.
The third rotational degree of freedom,
rotations around the rod symmetry axis, is only excited by frictional contacts
of particles in collisions. The ratio of the moments of inertia is
$J_{\perp}/J_{\parallel}\approx 70$, therefore such rotations need to be almost one
order of magnitude faster to reach equipartition.
For an estimate, we marked a few rods with dots to track their axial rotations. They
were evaluated during the cooling process, although
with much poorer statistics than for the other DoF. The related mean kinetic
energy turns out to be about one order of magnitude lower than those of the other DoF
\footnote{In molecular two-atomic gases at room temperature, such rotations about the axis with the smallest moment of inertia are not excited for quantum mechanical reasons.}.

\begin{figure}
   a)\includegraphics[width=\picwidth]{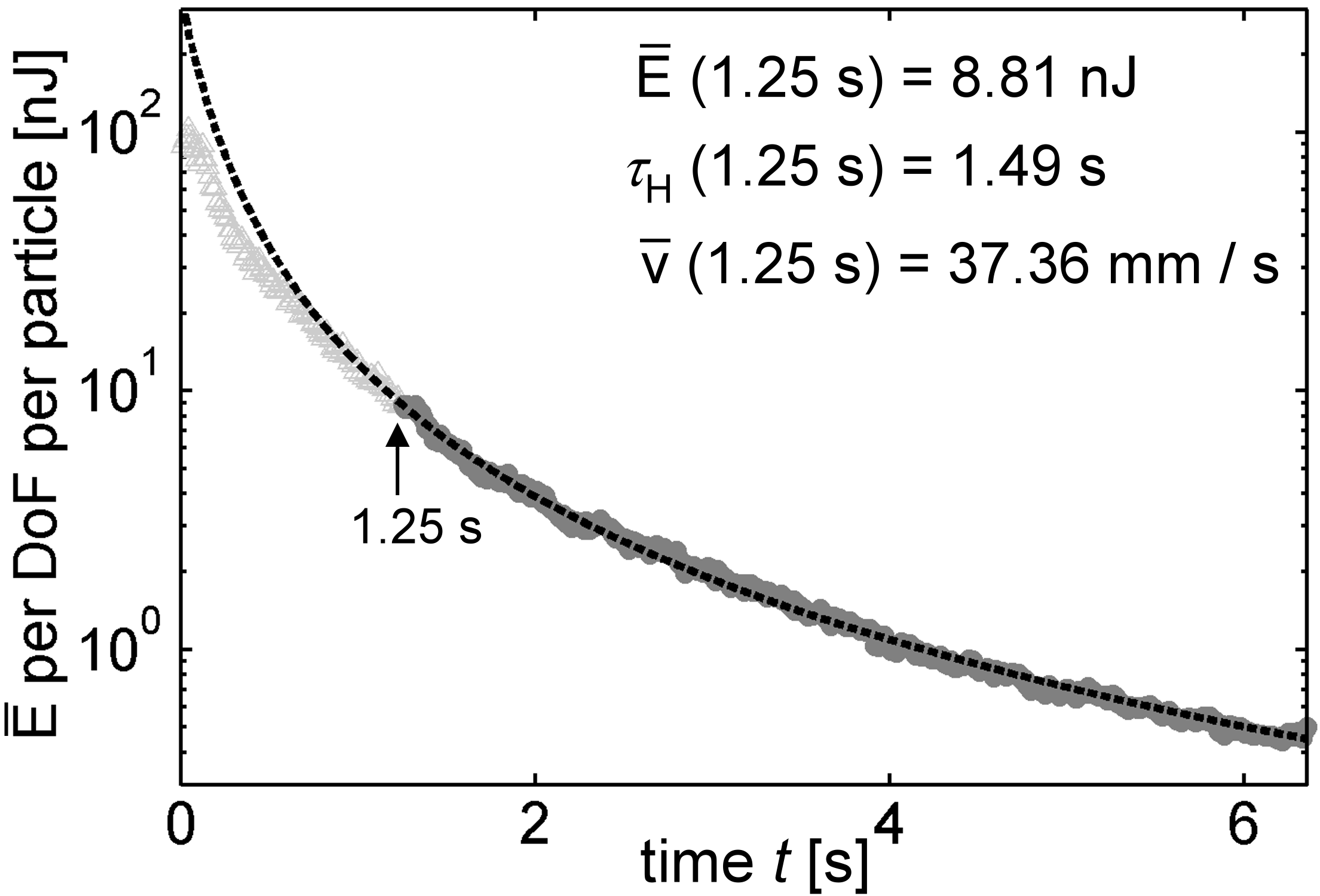}\\
   b)\includegraphics[width=\picwidth]{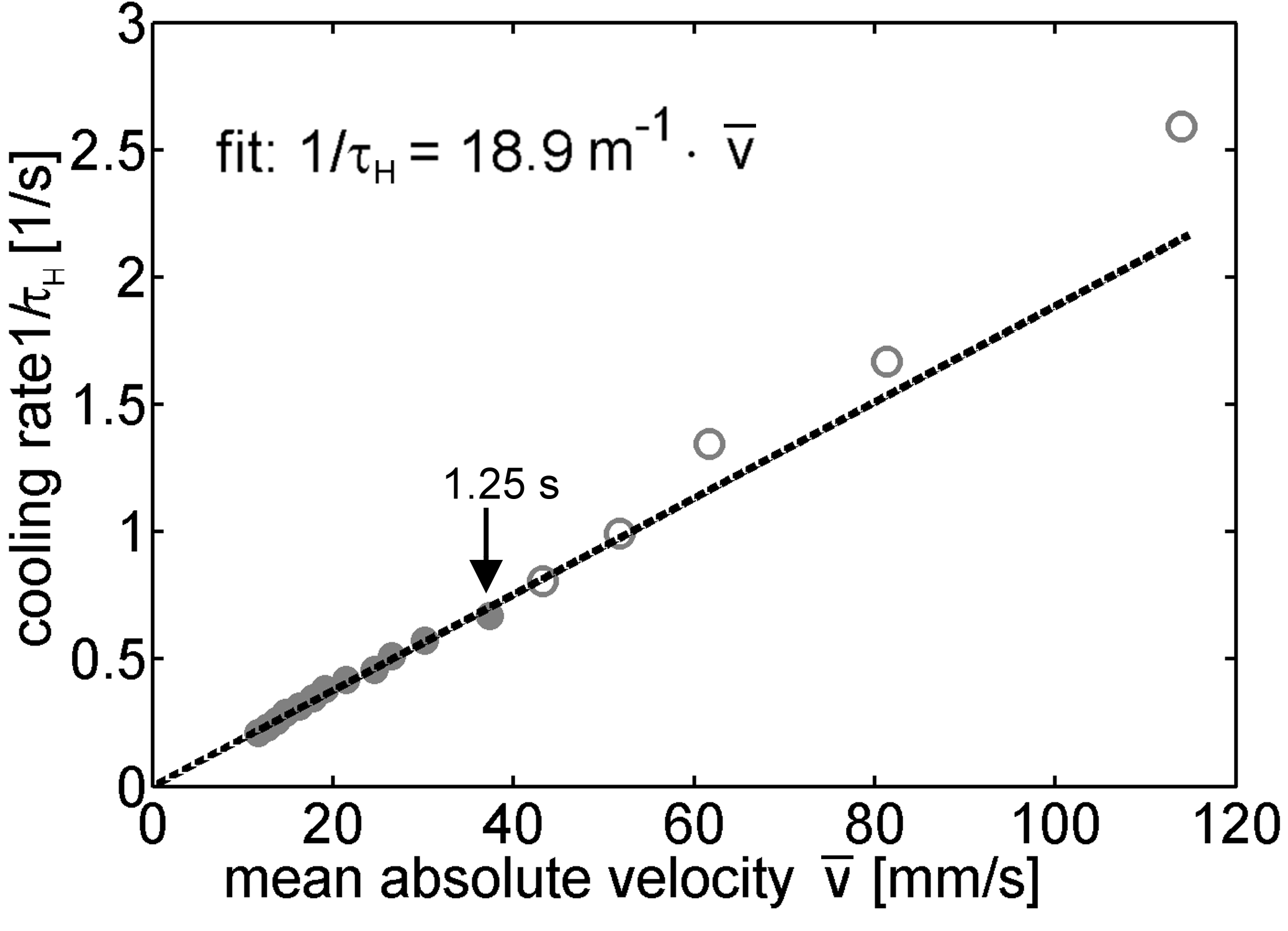}
  \caption{The granular cooling characteristics matches Haff's model:
  a) Decay of the average kinetic energy per degree of freedom and particle:
  after $\approx$ 1.5 s, the curve is in very good agreement with Haff's model \cite{Haff1983}, Eq. (\ref{eq:Haff}) (logarithmic fit). b) Comparison of the cooling rate and the measured mean velocities.
  The slope is $\xi(1-\varepsilon^2)/\lambda$. Solid symbols represent the data included in the fits, open symbols correspond to the initial 1.5 s when  Haff's equation is not yet applicable.
  }
  \label{Fig:3}
\end{figure}

\paragraph{Cooling:}
Haff \cite{Haff1983} predicted that the mean energy of a freely cooling granular
gas of {\em frictionless spheres} obeys the scaling
\begin{equation}
\label{eq:Haff}
 {E(t)}=\frac{E_0}{(1+t/\tau_H)^2},
\end{equation}
yielding $E(0)=E_0$ and $E(t)\propto t^{-2}$ for $t\gg \tau_H$, with the characteristic Haff time
 $\tau_H(E_0)$.
Haff starts with the dissipation rate
\begin{equation}\label{eq:Haff0}
\frac{\partial}{\partial t} \left(\frac {1}{2} \rho \bar v^2 \right) =
- \xi(1-\varepsilon^2)\rho\frac{\bar v^3}{s}
\end{equation}
Here, $\rho$ is the mass density of the system, $s$ is the mean grain separation (in our dilute system to be replaced by the mean free path $\lambda$), and $\varepsilon$ is the normal restitution coefficient,
$\bar v$ is the mean absolute velocity. The factor $\xi$ depends upon the dimensionality of the system, it basically accounts for the fact
that only the relative velocities of colliding particles (in our system also relative rotations) are relevant for the energy loss, which is then redistributed among all DoF.

Haff makes the assumption that the distance between the grains is small compared to their diameter. He sets the quantity $\overline {v^2}$, which is related to the average kinetic energy, equal to $\bar v^2$, the square of the mean absolute velocity, that determines the collision rate $\bar v/s$.
The Haff time describing cooling from a given instant $t_i$ is given by
\begin{equation}\label{eq:tauHaff}
\tau_H(t_i)=\lambda\,[\xi(1-\varepsilon^2) \bar v (t_i)]^{-1}.
\end{equation}
 Villemot and Talbot \cite{Villemot2012} have extended
this model to ellipsoidal particles, and Rubio-Largo et al. \cite{Rubio-Largo2016} considered ellipsoids
and spherocylinders, but aspect ratios in both studies were well below that of our rods. The corrections
concern the prefactor $\xi$ but leave the rest of the predictions unchanged.
Figure \ref{Fig:3}a shows that Eq.~(\ref{eq:Haff}) fits the experimental data of the mean total energy qualitatively very well after some initial period of $\approx 1.25$~s. Kanzaki et al.~\cite{Kanzaki2010}
predicted an exponent $-5/3$ of the decay of translational DoFs for viscoelastic ($\varepsilon(v)$) anisotropic grains in 2D from numerical simulations.
In clear contrast, our experiments are in excellent agreement with an exponent $-2$ for the long-term decay.

The initial discrepancy is easily understood: Immediately after excitation, the system is not spatially homogeneous, for example, particles are 'hotter' near the exciting plates. The
individual DoF are at very different granular temperatures. Therefore the Haff fit of the data after 1.25 s (dashed line in Fig.~\ref{Fig:3}a) overestimates
the initial $E(t)$. After about 1.25~s, the system is in the homogeneous cooling regime. This is confirmed by the analysis of the mutual dependence of $\tau_H$ and $\bar v$ (Fig.~\ref{Fig:3}b: The experimental data were fitted with Eq.~(\ref{eq:Haff}) starting at different initial
times $t_i>0$ during cooling, and the inverse of the fit parameter $\tau_H(t_i)$ was related to the momentary mean absolute velocities $\bar v(t_i)$. The linear fit confirms Eq.~(\ref{eq:tauHaff}) and yields the fit value
$\xi=18.9~\mbox{m}^{-1}\lambda /(1-\varepsilon^2) $.

\begin{figure}
  \includegraphics[width=\picwidth]{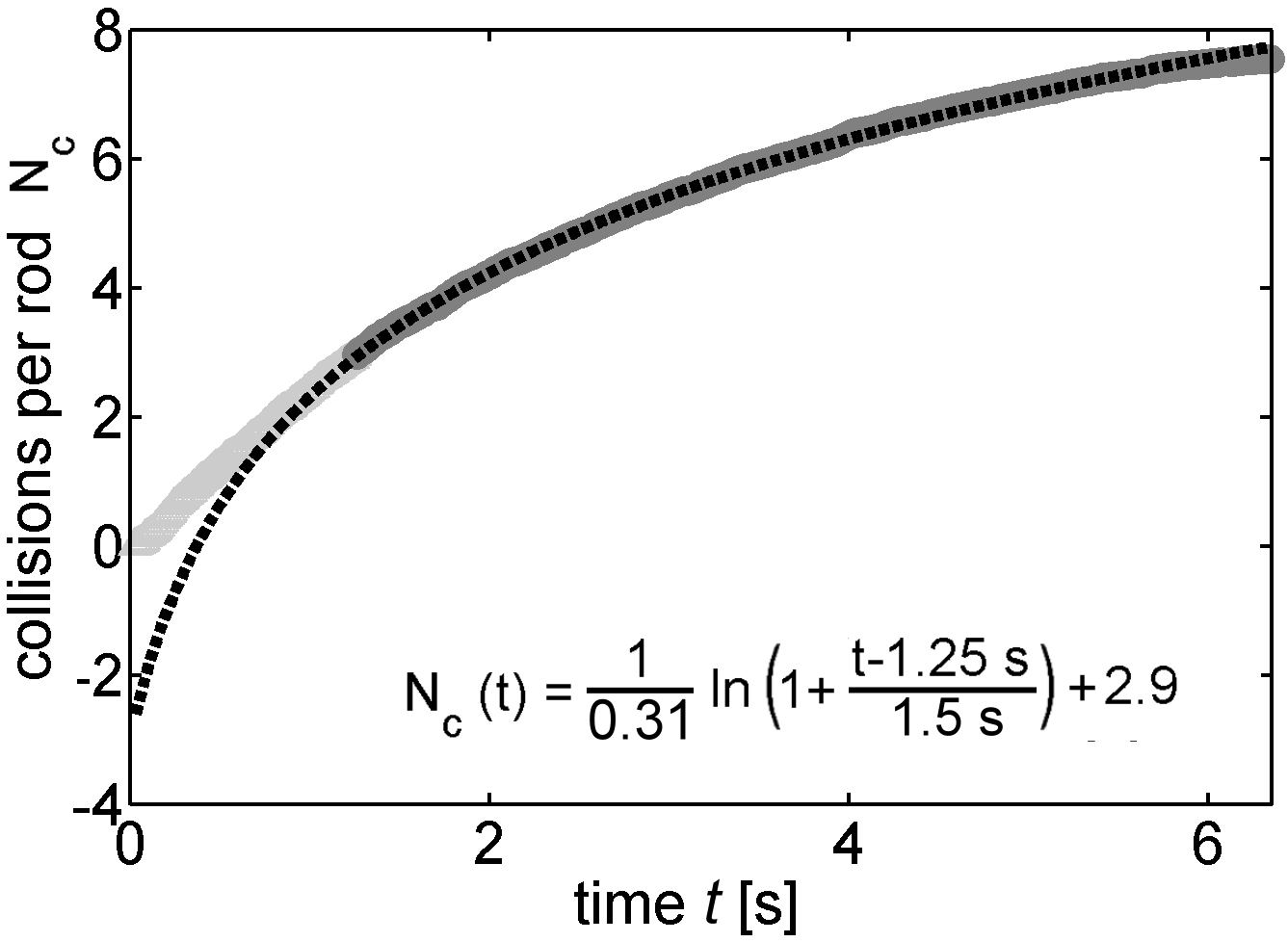}
  \caption{Model parameters are derived from the particle collision statistics:
  Cumulated number of collisions per particles determined from rate of collision of traced (colored)  particles in the videos. Only the dark data points were fitted.
  }
  \label{Fig:4}
\end{figure}

A further test of the model is the experimental determination of the cumulated collision number
per particle
$$N_c = \frac{1}{\xi(1-\varepsilon^2)}\ln\left(1+\frac{t}{\tau_H}\right) +\mbox{const}$$
(Figure \ref{Fig:4}).
The dashed line is a fit through the data for $t>1.25$~s. Again, the initial values deviate from that fit,
partially because the homogeneous cooling is not yet reached, and partially because the particles are very fast so that not all collisions may have been detected unambiguously in the videos (both reflected in the offset). Together with the fit in Fig.~\ref{Fig:3}, this yields a value $\lambda = 0.31/18.9$~m $\approx 1.64$~cm, in excellent agreement with our geometrical estimate.

In our experiment, the following complications affect the interpretation of the factor $\xi$:
In Haff's model, the energy loss per collision is distributed among three DoF. In our system, translational
energy is partially converted in rotational energy and vice versa
\cite{Luding1998,Rubio-Largo2016}, and the loss of rotational energy per
collision is much more difficult to estimate than for frictionless spheres. In addition, the kinetic energy share of rotations about the long rod axis is known only to the order of magnitude.
For an estimation of $\tau_H$ with Eq.~(\ref{eq:tauHaff}), $\varepsilon$ and $\bar v$ can in principle be determined experimentally, but the factor $\xi$ is unknown.
Numerical simulations \cite{Villemot2012,Rubio-Largo2016} may give some hints,
but it is not clear whether these results can be extrapolated
to our particles with aspect ratios above 7, and whether the results of Ref. \cite{Villemot2012} can be applied to rodlike particles at all.
An additional problem is that we find a preferential alignment of the rods in flight direction (see below), so that the collision statistics is not the same as for completely random rod orientations.

The velocity distributions of the components $v_x,~v_y,~v_z$ are non-Gaussian. Within our statistical accuracy, their shape remains nearly unchanged during the homogeneous cooling process.
The kurtosis has no trend in our experimental data during homogeneous cooling, it fluctuates between 3 and 4, with an average of 3.5.
The measured ratio of $\overline{v^2}$ and $\overline v^2$ was between 1.2 and 1.32, slightly increasing with time.
This is in clear contrast to  Gaussian velocity distributions reported in simulations
\cite{Villemot2012,Rubio-Largo2016}.

\begin{figure}[ht]
{\includegraphics[width=\columnwidth]{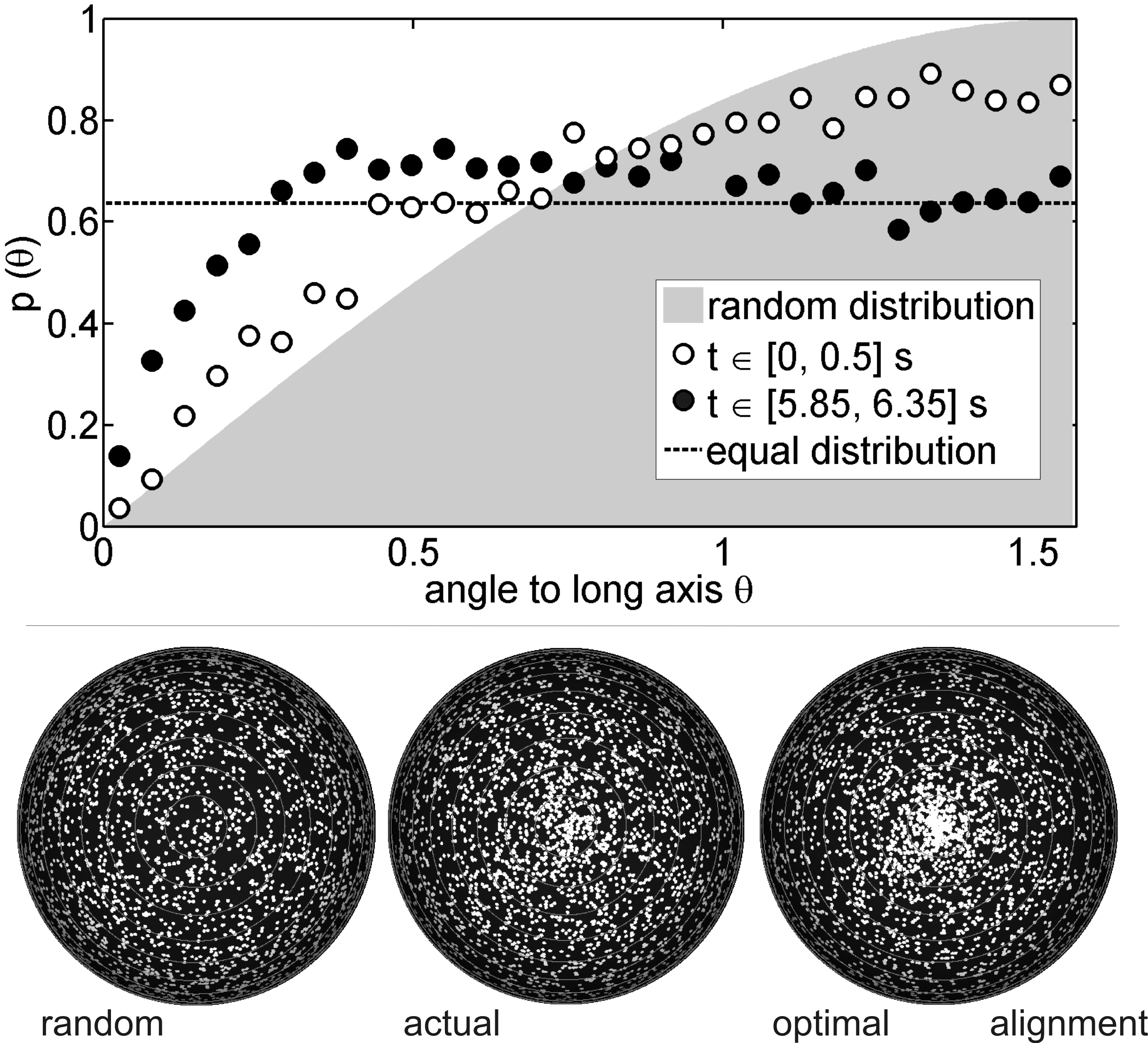}}
\caption{Rods slightly align respective to the momentary flight direction:
angle between the rod symmetry axis and $\vec{v}$ during the initial
$0.5~\rm s$ of cooling, and during the last 0.5 s of the experiment.
The pictures below illustrate the distribution $p(\theta)$ by bright dots projected on a unit sphere,
looking in flight direction, left: isotropically distributed rod axes,
middle: experiment at $t\approx 6$~s, right: best possible alignment.}
\label{Fig:5}
\end{figure}

\paragraph{Alignment:}
One particular feature of anisotropic grains is their tendency to
align in shear flow \cite{BorzsonyiReview}, or in active matter \cite{Bricard2013}.
Correlations between velocity and
orientation were reported in simulations of hard needles~\cite{Huthmann1999}.
We find similar correlations in the distribution of relative rod orientations $p(\theta)$,
where $\theta=0^{\circ}$ resembles spear-like orientation. Immediately after excitation, the alignment angles
are distributed almost isotropically ($p(\theta)\propto \sin\theta$), this can be attributed to the random excitation by the vibrating container walls. During cooling, rods are more often in spear-like orientation, the probability of angles $\theta<45^{\circ}$ increases, while it decreases for $\theta>45^{\circ}$ (Fig.~\ref{Fig:5}).
This can be understood intuitively: rod-rod collisions are more probable when the rod axis is perpendicular to the flight path.
The consequence is a slightly larger $\lambda$ than estimated for random rod orientations.
The effect is small, though, as illustrated in the bottom images of Fig.~\ref{Fig:5}.
Note that the highest possible alignment of rotating rods would be an equal distribution
$p(\theta)=2/\pi$ (black line in the image), when all rotations occur about short axes perpendicular to the flight path (like spokes of a wheel).

\section{Conclusions and summary}
The experimental study of homogeneous free cooling of a 3D granular gas in micro-gravity
demonstrates that Haff's scaling law of the energy loss with time is surprisingly robust, even though several central assumptions
are not fulfilled, e.g. friction and shape-anisometry of the grains, and non-Gaussian velocity distributions even in the homogeneous cooling state. Energies become nearly equally distributed among the DoF in the
homogeneously cooling state, with a slight excess ($\approx 10..20$~\%)
in the translational DoF. Even the purely friction-coupled rotations around the long axis are excited, 
albeit with one order of magnitude lower mean energy.
A gradual alignment of rods in flight direction is also documented, explainable by the lower collision probability in this flight orientation.

The detailed mechanisms underlying the collective dynamics of this system
e.g. the exact role of particle shape, contact parameters, confinement and spatial inhomogeneities,
are still to be explored. The present data may serve as benchmarks for computational studies.
In perspective, only the combined efforts of theoretical work allowing for conditions realizable in experiments, 
and quantitative experiments with materials
of different shapes, friction and elastic properties, may bridge the gap from individual grain collision to ensemble dynamics of granular gases in an extendable and realistic manner.

\begin{acknowledgments}
The authors acknowledge funding by the German Aerospace Center DLR,
projects 50WM1241 and 50WM1344, and by the German Science Foundation (DFG)
grant STA 425/34-1. We cordially thank the ZARM staff in Bremen for their
excellent support in the drop tower experiments.
\end{acknowledgments}

\section{Methods}
The setup consisting of a box with two laterally movable side walls, illumination and two cameras recording perpendicular perspectives (similar to Ref.~\cite{Harth2013}),
see Fig.~\ref{Fig:ExperimentBasics} is integrated into a ZARM catapult capsule. When the capsule is released from the catapult, the rods lift off from the base plate.
We prepare an initial excited state by injecting energy through vibrations of the lateral walls in $x$-direction ($3~\rm mm$ amplitude, 30 Hz) for $1.5\dots 2~\rm s$. The velocity distributions in this
initial state
are spatially inhomogeneous
(see, e. g. \cite{Harth2015,Yanpei2011}), while the spatial distribution of rod positions and orientations remain almost homogeneous throughout the complete experiment, see Fig.~\ref{Fig:ExperimentBasics}c and supplementary movie.
After the excitation is stopped,
3D trajectories of the colored rods in the ensembles are tracked in the two perspective views. Black rods
provide the thermal background, it is technically too demanding to distinguish and track more than 50...60 rods in
the ensemble. Because particles are sometimes obscured by others, individual trajectories cannot always be followed continuously, see Fig.~\ref{Fig:ExperimentBasics}. Partially necessary manual tracking is the most time-consuming task of the data analysis.
We mark all collisions in the trajectories and reorientations
manually and fit segments of the translational motion between collisions linearly to reduce sampling noise. Angular velocities are computed from the vector
product of the rod orientations in consecutive time steps, and their absolute values are averaged in the intervals between collisions for the same reason. Only rotations about the short rod axes are
analyzed systematically. The statistical analysis combines data from 15 independent experimental runs,
providing $\approx 130 000$ data points ($\approx 200$ data points for each 10 ms time step).

The coefficients of restitution of our particles are not known. A certain estimate of the normal coefficient of restitution was obtained from collisions with a container wall \cite{Harth2013}, the total loss of kinetic energy per collision yields $\varepsilon\approx0.54$. This is not necessarily a good estimate
for particle-particle collisions, which are very difficult to evaluate.

\bibstyle{prbst}
\bibliography{GAGa}

\end{document}